\documentclass[aps,pra,a4paper,amsmath,amssymb,tightenlines,twocolumn,superscriptaddress,showpacs,nofootinbib]{revtex4}

\usepackage{amsmath}

  \newcommand{\Bkt}[1]{\left( #1 \right)}
  \newcommand{\bkt}[2][]{#1( #2 #1)}
  \newcommand{\Bktcl}[1]{\left\{ #1 \right\}}
  \newcommand{\bktcl}[2][]{#1\{ #2 #1\}}
  \newcommand{\Bktsq}[1]{\left[ #1 \right]}
  \newcommand{\bktsq}[2][]{#1[ #2 #1]}

  \newcommand{\norm}[2][]{#1 \lVert #2 #1 \rVert}
  \newcommand{\Abs}[1]{\left \lvert #1 \right \rvert}
  \newcommand{\abs}[2][]{#1 \lvert #2 #1 \rvert}

  \renewcommand{\vec}[1]{\mathbf{#1}}
  
  \newcommand{\del}[0]{\nabla}

  \newcommand{\Exval}[1]{\left\langle #1 \right\rangle}
  \newcommand{\exval}[2][]{#1\langle #2 #1\rangle}
  \DeclareMathOperator{\Tr}{Tr}

\usepackage{graphicx}
\usepackage{rotating} 
\usepackage{psfrag}

\renewcommand{\v}{\vec}

\newcommand{\rC}{\text{\bf{C}}} 
\newcommand{\rI}{\text{\bf{I}}} 
\newcommand{\PC}{\mathcal{P}_{\rC}}
\newcommand{\bef}{\hat{\psi}}
\newcommand{\bfI}{\bef_{\rI}}
\newcommand{\cf}{\psi_{\rC}}

\newcommand{\CF}{c-field}

\newcommand{\Tkt}{T_\text{KT}}
\newcommand{\md}{\hat{\v{p}}} 
  
\newcommand{\punpair}{p_u} 

\begin{document}

\title{Vortex pairing in two-dimensional Bose gases}
\author{Christopher J. Foster}
\affiliation{The University of Queensland, School of Mathematics and Physics,
ARC Centre of Excellence for Quantum-Atom Optics, Qld 4072, Australia}
\author{P. Blair Blakie}
\affiliation{The Jack Dodd Centre for Quantum Technology, Department of
Physics, University of Otago, Dunedin, New Zealand.}
\author{Matthew J. Davis}
\affiliation{The University of Queensland, School of Mathematics and Physics,
ARC Centre of Excellence for Quantum-Atom Optics, Qld 4072, Australia}

\begin{abstract}
  Recent experiments on ultracold Bose gases in two dimensions have provided
  evidence for the existence of the Berezinskii-Kosterlitz-Thouless (BKT)
  phase via analysis of the interference between two independent systems.
  In this work we study the two-dimensional quantum degenerate Bose gas at
  finite temperature using the  projected 
  Gross-Pitaevskii equation classical field method.  While this describes the highly occupied modes of
  the gas below a momentum cutoff, we have developed a method to incorporate
  the higher momentum states in our model.  We concentrate on finite-sized
  homogeneous systems in order to simplify the analysis of the vortex pairing.
  We determine the dependence of the condensate fraction on temperature and
  compare this to the calculated superfluid fraction.  By measuring the first
  order correlation function we determine the boundary of the Bose-Einstein condensate and BKT
  phases, and find it is consistent with the superfluid fraction decreasing to
  zero.  We reveal the characteristic unbinding of vortex pairs above the BKT
  transition via a coarse-graining procedure.  Finally, we model the procedure used in experiments to infer system correlations  [Hadzibabic et al., Nature \textbf{441}, 1118 (2006)], and quantify its level of agreement with directly calculated \textit{in situ} correlation functions. 
\end{abstract}


\maketitle

\section{Introduction}
At low temperatures a three-dimensional (3D) Bose gas can undergo a phase 
transition to a Bose-Einstein condensate. In contrast, thermal fluctuations 
prevent a two-dimensional (2D)  Bose gas from making a phase transition to an 
ordered state, in accordance with the Mermin-Wagner-Hohenberg theorem 
\cite{Mermin1966,Hohenberg1967}. However, the 2D Bose gas supports topological
defects in the form of vortices, and in the presence of interactions can
instead undergo a Berezinskii-Kosterlitz-Thouless (BKT)  
\cite{Berezinskii1971,Kosterlitz1973,Posazhennikova2006} transition to a 
quasi-coherent superfluid state. The BKT transition was first observed in 
liquid helium thin films \cite{Bishop1978}, however, more recently, evidence for 
this transition has been found in dilute Bose gases 
\cite{Stock2005,Hadzibabic2006,Kruger2007,Schweikhard2007a,Clade2008} (also see \cite{Gorlitz2001,Smith2004}).

Ultracold gases have proven to be beautiful systems for making direct comparisons between experiment and \textit{ab initio} theory. Experiments in the 2D regime present a new challenge for theory as  strong fluctuations    invalidate mean-field theories (e.g.\ see \cite{Prokofev2001,Prokofev2002,Posazhennikova2006,Bisset2009d,Mathey2009a,Sato2009a,Gies2004a,Schumayer2007a}), and only recently have quantum Monte Carlo \cite{Holzmann2008,Holzmann2009b} and classical field (c-field)  \cite{Simula2006,Simula2008,Bisset2009} methods been developed that are directly applicable to the experimental regime. 
 
In this paper we study a uniform Bose gas of finite spatial extent and parameters corresponding to current experiments. To analyze this system we use the projected Gross-Pitaevskii equation (PGPE), a c-field technique suited to studying finite temperature Bose fields with many highly occupied modes. 
We develop a technique for extracting the superfluid density based on linear response properties, and use this to understand the relationship between superfluidity and condensation in the finite system.   

With this formalism we then examine two important applications: 
First, we provide a quantitative validation of the interference technique used in the ENS experiment to determine the nature of two-point correlation in the system. To do this we simulate the interference pattern generated by allowing two independent 2D systems to expand and interfere. Then applying the experimental fitting procedure to analyze the interference pattern we can extract the inferred two point correlations, which we can then compare against the \textit{in situ} correlations that we calculate directly. 
Second, we examine the correlations between vortices and antivortices in the system to directly quantify the emergence of vortex-antivortex pairing in the low temperature phase. A similar study was made by Giorgetti \textit{et al.} using a semiclassical field technique \cite{Giorgetti2007}. We find results for vortex number and vortex pair distributions consistent with their results, and we show how a coarse graining procedure can be used to reveal the unpaired vortices in the system. 
 
We now briefly outline the structure of this paper: In Sec.~\ref{Sform} we review the 2D Bose gas and relevant BKT physics. In Sec.~\ref{sec:method} we outline the c-field technique and how it is specialized to describing a uniform, but finite, 2D  Bose gas. In Sec.~\ref{sec:res} we present our main results, before concluding.

\section{Formalism}\label{Sform}
Here we consider a dilute 2D Bose gas described by the Hamiltonian
\begin{equation}
\hat{H}=\int d^{2}\mathbf{x}\,\hat{\psi}^{\dagger}\left\{-\frac{\hbar^2\nabla^2_{\mathbf{x}}}{2m}\right\}\hat{\psi}+\frac{\hbar^2g}{2m}\int d^{2}\mathbf{x}\,\hat{\psi}^\dagger\hat{\psi}^{\dagger}\hat{\psi}\hat{\psi},\end{equation}
where $m$ is the atomic mass, $\mathbf{x}=(x,y)$, and $\hat{\psi}=\hat{\psi}(\mathbf{x})$ is the bosonic field operator.

We take the two-dimensional geometry to be realized by tight confinement in the $z$ direction that restricts atomic occupation to the lowest  $z$ mode. 
The dimensionless 2D coupling constant is 
\begin{equation}
{g}=\frac{\sqrt{8\pi}a}{a_{z}},
\end{equation}
with $a_{z}$ the spatial extent of the $z$ mode  \footnote{For example, for tight harmonic confinement of frequency $\omega_z$ we have $a_z=\sqrt{\hbar/m\omega_z}$.} and $a$ the s-wave scattering length. We will assume that $a_{z}\gg a$
so that the scattering is approximately three-dimensional \cite{Petrov2000},
a condition well-satisfied in the ENS and NIST experiments 
\cite{Stock2005,Hadzibabic2006,Kruger2007,Clade2008}.
For reference, the ENS experiment reported in \cite{Hadzibabic2006} had ${g}\approx0.15$,  whereas
in the NIST experiments ${g}\approx0.02$ \cite{Clade2008}.

In contrast to experiments we focus here on the uniform case; no trapping 
potential in the $xy$ plane is considered.  We perform finite sized calculations
corresponding to a square system of size $L$ with periodic boundary conditions.
Working in the finite size regime simplifies the simulations and is more 
representative of current experiments.  We note that the thermodynamic limit
corresponds to taking $L\to\infty$ while keeping the density,
$n=\langle\hat{\psi}^\dagger\hat{\psi}\rangle$, constant.

\subsection{Review of BKT physics}
The BKT superfluid phase has several distinctive characteristics, which we briefly review.
\subsubsection{First order correlations}
Below the BKT transition the first-order correlations decay according to an inverse power law:
\begin{equation} \label{eqn:algebraic_decay}
g^{(1)}(\vec{x},\vec{x}') \propto \norm{\vec{x}-\vec{x}'}^{-\alpha}.
\end{equation}
Systems displaying such \emph{algebraic decay} are said to exhibit 
\emph{quasi-long-range order} \cite{Chaikin_Lubensky1995}.
This is in contrast to both the high temperature (disordered phase) in 
which the correlations decay exponentially, and long-range ordered case of the 3D Bose gas in which $g^{(1)}\to\rm{const.}$ for $\norm{\vec{x}-\vec{x}'}\to\infty$.


\subsubsection{Superfluid density}
Nelson and Kosterlitz \cite{Nelson1977} found that the exponent of the
algebraic decay is related to the ratio of the superfluid density and
temperature.  To within logarithmic corrections
\begin{equation} \label{eqn:alpha_rhos_T}
	\alpha(T) = \frac{1}{\lambda^2\rho_s(T)},
\end{equation}
where $\rho_s$ is the superfluid density and $\lambda=h/\sqrt{2\pi mk_BT}$ is 
the thermal de Broglie wavelength.  Furthermore, Nelson and 
Kosterlitz showed that this ratio converges to a universal constant as
the transition temperature, $\Tkt$, is approached from below: $\lim_{T\to \Tkt^-}
\alpha(T) = 1/4$ (i.e.,~$\rho_s\lambda^2=4$).  Thus, the superfluid fraction undergoes a universal jump from  
$\rho_s(\Tkt^+)=0$ to $\rho_s(\Tkt^-)=4/\lambda^2$ as the temperature decreases through $\Tkt$.

\subsubsection{Vortex binding transition}
Another important indicator of the BKT transition is the behavior of
topological excitations, which are quantized vortices and antivortices in the
case of a Bose gas.  A single vortex has energy which scales with the logarithm of the
system size.  At low temperatures this means that the free energy for a single
vortex is infinite (in the thermodynamic limit), and vortices cannot exist in isolation.  As originally
argued in \cite{Kosterlitz1973}, the entropic contribution to the free energy
also scales logarithmically with the system size, and will dominate the free energy at high temperatures allowing unbound vortices to proliferate.  This argument
provides a simple estimate for the BKT transition temperature.

Although unbound vortices are thermodynamically unfavored at $T<\Tkt$, bound pairs of
counter-rotating vortices may exist since the total energy of such a pair is
finite \footnote{The vortex-antivortex pair energy depend on the pair size 
rather than the system size.}.  This leads to a distinctive qualitative
characterization of the BKT transition: as the temperature increases through
$\Tkt$ pairs of vortices unbind.

%
%



\subsubsection{Location of the BKT transition in the dilute Bose gas}
While the relation $\rho_s(\Tkt^-)=4/\lambda^2$ between the superfluid density and temperature at the transition is universal, the total density, $n$, at the transition is not.
General arguments \cite{Popov,Kagan1987,Fisher1988} suggest that the transition point for the dilute uniform 2D Bose gas is given by
\begin{equation}
(n\lambda^2)_\text{KT} = \ln\left(\frac{\xi}{{g}}\right), \label{critpsd}
\end{equation}
where  $\xi$ is a constant.
Prokof\'ev, Ruebenacker and Svistunov \cite{Prokofev2001, Prokofev2002} studied
the homogeneous Bose gas using Monte Carlo simulations of an equivalent
classical $\phi^4$ model on a lattice.  Using an extrapolation to the
infinite-sized system, they computed a value for the dimensionless constant,
$\xi = 380 \pm 3$.  By inverting Eq.~(\ref{critpsd}), we obtain the  BKT critical temperature for the infinite
system 
\begin{equation}
	\Tkt^\infty = \frac{2\pi \hbar^2 n}{m k_B \ln \Bkt{\xi \hbar^2 / m {g}}}.
\end{equation} 
We use the superscript $\infty$ to indicate that this result holds in the thermodynamic limit.

\section{Method}
\label{sec:method}

\subsection{c-field and incoherent regions} \label{secPGPEformalism}
We briefly outline the PGPE formalism, which is developed in detail in Ref. \cite{cfieldRev2008}.  
The Bose field operator is split into two parts according to
\begin{equation}
\hat\psi(\mathbf{x}) = \cf(\mathbf{x}) + \bfI(\mathbf{x}),\label{EqfieldOp}
\end{equation}
where $\cf$ is the coherent region \CF\  and $\bfI$ is the incoherent field operator (see \cite{cfieldRev2008}).
These fields are defined as the low and high energy projections of the full 
quantum field operator, separated by the cutoff wave vector $K$. In our theory 
this cutoff is implemented in terms of the plane wave eigenstates 
$\{\varphi_\v{n}(\mathbf{x})\}$ of the time-independent single particle 
Hamiltonian, that is,
\begin{align}
\varphi_\v{n}(\mathbf{x}) &= \frac{1}{{L}}e^{-i\mathbf{k}_\v{n}\cdot\mathbf{x}}, \\
\mathbf{k}_\v{n} &= \frac{\pi}{L}\v{n},
\end{align}
with $\v{n} = (n_x, n_y) \in\mathbb{Z}^2$.
The fields are thus defined by
\begin{align}
\cf(\mathbf{x}) &\equiv \sum_{\v{n}\in\rC}c_\v{n}\varphi_\v{n}(\mathbf{x}),\label{eqn:Cfield}\\
\bfI(\mathbf{x}) &\equiv \sum_{\v{n}\in\rI}\hat{a}_\v{n}\varphi_\v{n}(\mathbf{x}),
\end{align}
where the $\hat{a}_\v{n}$ are Bose annihilation operators, the $c_\v{n}$ are complex amplitudes, and the sets of quantum numbers defining the regions are 
\begin{align}
\rC &= \{\v{n}:\norm{\mathbf{k}_\v{n}}\le K\},\\
\rI &= \{\v{n}:\norm{\mathbf{k}_\v{n}}> K\}. 
\end{align}

\subsubsection{Choice of $\rC$ region}\label{cregionchoice}
In general, the applicability of the PGPE approach to describing the finite temperature gas relies on an appropriate choice for $K$, so that the modes at the cutoff have an average occupation of order unity. In this work we choose an average of
five or more atoms per mode using a procedure discussed in appendix \ref{sec:sim_details}.  This choice means that all the modes in $\rC$ are appreciably occupied, justifying the classical field replacement $\hat{a}_\v{n}\to c_\v{n}$. In contrast the $\rI$ region contains many sparsely occupied modes that are particle-like and would be poorly described using a classical field approximation. 
Because our 2D system is critical over a wide temperature range, additional care is needed in choosing $\rC$. Typically strong fluctuations occur in the infrared modes up to the energy scale  $\hbar^2gn/m$. Above this energy scale the modes are well described by mean-field theory (e.g.\ see the discussion in \cite{Kashurnikov2001a,Prokofev2001}). 
For the results we present here, we have
\begin{equation}
\frac{\hbar^2K^2}{2m} \gtrsim \frac{\hbar^2g}{m}n\label{validitycond}
\end{equation}
for simulations around the transition region and at high temperature.  At
temperatures well below $\Tkt$, the requirement of large modal occupation near
the cutoff competes with this condition and we favor the former at the expense
of violating Eq.~\eqref{validitycond}.

\subsubsection{PGPE treatment of $\rC$ region}\label{SecformalismPGPE}

The equation of motion for $\cf$ is the PGPE
\begin{equation}
i\hbar\frac{\partial \cf }{\partial t} = -\frac{\hbar^2\nabla^2_{\mathbf{x}}}{2m}\cf + \frac{\hbar^2g}{m} \PC\left\{ \abs{\cf}^2\cf\right\}, \label{PGPE}
\end{equation}
where the projection operator 
\begin{equation}
\PC\{ F(\mathbf{x})\}\equiv\sum_{\v{n}\in\rC}\varphi_{\v{n}}(\mathbf{x})\int
d^2\mathbf{x}'\,\varphi_{\v{n}}^{*}(\mathbf{x}') F(\mathbf{x}'),\label{eq:projectorC}\\
\end{equation}
formalizes our basis set restriction of $\cf$ to the $\rC$ region. The main approximation used to arrive at the PGPE is to neglect dynamical couplings to the incoherent region \cite{Davis2001b}.

We assume that Eq.~(\ref{PGPE}) is ergodic \cite{Davis2001a}, so that the 
microstates \{$\cf$\} generated through time evolution form an unbiased 
 sample of
the equilibrium microstates.  Time averaging can then be used to obtain
macroscopic equilibrium properties.  We generate the time evolution by solving
the PGPE with three adjustable parameters: (i) the  cutoff wave vector, $K$,
that defines the division between $\rC$ and $\rI$, and hence the number of
modes in the $\rC$ region; (ii) the number of $\rC$ region atoms, $N_{\rC}$;
(iii) the total energy of the $\rC$ region, $E_\rC$. The last two quantities,
defined as 
\begin{align}
E_{\rC} &= \int d^2\mathbf{x}\,\cf^*\left(-\frac{\hbar^2\nabla^2_{\mathbf{x}}}{2m} +  \frac{\hbar^2g}{2m} \abs{\cf}^2\right)\cf,\label{Ec}\\
N_{\rC} &= \int d^2\mathbf{x}\,\abs{\cf(\mathbf{x})}^2,
\end{align}
are important because they represent constants of motion of the PGPE
\eqref{PGPE}, and thus control the thermodynamic equilibrium state of the system.

\subsubsection{Obtaining equilibrium properties for the $\rC$ region}\label{PGPEeqprops}
To characterize the equilibrium state in the $\rC$ region it is necessary to
determine the average density, temperature and chemical potential, which in 
turn allow us to characterize the $\rI$ region (see Sec.~\ref{sec:above_cutoff}).  
These and other $\rC$ region quantities can be computed by time-averaging, e.g,
the average $\rC$ region density is given by
\begin{equation}
n_{\rC}(\mathbf{x})  \approx \frac{1}{M_s}\sum_{j=1}^{M_s}\Abs{\cf(\mathbf{x},t_j)}^2,\label{nc2}
\end{equation}
where $\{t_j\}$ is a set of $M_s$ times (after the system has been allowed to
relax to equilibrium) at which the field is sampled. We typically use 2000
samples from our simulation to perform such averages over a time of $\sim 16$ s.
Another quantity of interest here is the first order correlation function,
which we calculate directly via the expression
\begin{equation}
G^{(1)}_{\rC}(\v{x},\v{x}')  \approx \frac{1}{M_s}\sum_{j=1}^{M_s}\cf^*(\mathbf{x},t_j)\cf(\mathbf{x}',t_j).\label{Gc1}
\end{equation}

Derivatives of entropy, such as the temperature ($T$) and chemical potential ($\mu_{\rC}$) can be calculated by time averaging appropriate quantities constructed from the Hamiltonian (\ref{Ec}) using the Rugh approach \cite{Rugh1997a}.  The detailed implementation of the Rugh formalism for the PGPE is rather technical and we refer the reader to Refs. \cite{Davis2003,Davis2005} for additional details of this procedure.

A major extension to the formalism of the PGPE made in this work is the development of a method for extracting the superfluid fraction, $\rho_s$, from these calculations. 
For this we use linear response theory to relate the superfluid fraction to the long wavelength limit of the second order momentum density correlations.
An extensive discussion of this approach, and the numerical methods used to implement it, are presented in appendix \ref{sec:superfluid_fraction}.

\subsection{Mean-field treatment of $\rI$ region} \label{sec:above_cutoff}
Occupation of the $\rI$ region modes, $N_\rI$, accounts for about 25\% of the total
number of atoms at temperatures near the phase transition.  We assume a time
independent state for the $\rI$ region atoms defined by a Wigner function
\cite{Naraschewski1999}, allowing us to calculate quantities of interest by
integrating over the above-cutoff momenta, $k > K$
\cite{Bezett2008, Davis2006}.

Our assumed Wigner function corresponds to the self-consistent Hartree-Fock
theory as applied in \cite{Davis2006}.  In two dimensions this is
\begin{equation}\label{eqn:wignerFxn}
W_\rI(\v{k},\v{x})
  = \frac{1}{(2\pi)^2} \frac{1}{e^{(E_\text{HF}(\v{k}) - \mu)/k_B T} - 1 },
\end{equation}
where 
\begin{equation}
E_\text{HF}(\v{k}) =\frac{\hbar^2\v{k}^2}{2m} + \frac{2\hbar^2g}{m}(n_\rC + n_\rI),\label{EHF}
\end{equation}
is the Hartree-Fock energy,  $n_\rI$ is the $\rI$ region density, and $\mu = \mu_\rC + 2\hbar^2gn_\rI/m$ is the chemical potential (shifted by the mean-field interaction with the $\rI$ region atoms).  Note that the average densities are constant in the uniform system, so $W_\rI(\v{k},\v{x})$ has no explicit $\v{x}$ dependence, however, we include this variable for generality when defining the associated correlation function. 

The $\rI$ region density appearing  in Eq.~\eqref{EHF} is given by
\begin{equation}
n_\rI= \int_{\norm{\v{k}} \ge  K} d^2\v{k}\, W_\rI(\v{k},\v{x}),
\end{equation}
with corresponding atom number $N_\rI=n_\rI L^2$; total number is simply
\begin{equation}
N=N_\rC+N_\rI.
\end{equation}
An analytic expression for $n_\rI$ and simplified procedure for numerically calculating the first order correlation function of the $\rI$ region atoms, $G^{(1)}_\rI$, can be obtained  by taking integrals over the phase space. These results are discussed in appendix \ref{sec:above_cutoff_integrals}. 

\subsection{Equilibrium configurations with fixed $T$ and $N$} \label{sec:initial_conditions}

Generating equilibrium classical fields with given values of $E_\rC$ and $N_\rC$
is straightforward since the PGPE simulates a microcanonical system
(see appendix \ref{sec:initial_given_Ec_Nc}).  However, we wish to
simulate systems with a given temperature and total number.  As described in
the preceding two sections these can only be determined after a simulation has
been performed.  In appendix \ref{sec:sim_details} we outline a procedure for
estimating values of $E_\rC$ and $N_\rC$ for desired values of $N$ and $T$
based on a root finding scheme using a Hartree-Fock-Bogoliubov analysis for the
initial guess.

\section{Results}\label{sec:res}

We choose simulation parameters in analogy with the Paris experiment of
Hadzibabic \textit{et al.}~\cite{Hadzibabic2006}.  This experiment used an
elongated atomic cloud of approximately $10^5$ $^{87}$Rb atoms, with a spatial extent (Thomas-Fermi lengths) of 120 $\mu$m and 10 $\mu$m along the two loosely
trapped $x$ and $y$ directions. The tight confinement in the $z$ direction was provided by an optical lattice.

Although our simulation is for a uniform system, we have chosen 
similar parameters where possible.  Our primary simulations are for a system in a square box with $L=100$ $\mu$m, with  $4{\times}10^5$ $^{87}$Rb atoms. We also present results for systems with $L=50$ $\mu$m and $L=200$ $\mu$m at the same density in order
to better understand finite-size effects. All simulations are for the case of
$g=0.15$ corresponding to the experimental parameters reported in \cite{Hadzibabic2006}.

The cutoff wave vector $K$ varied with temperature to ensure appropriate occupation of the highest modes (see Sec.~\ref{cregionchoice}). For the 100 $\mu$m system, the number of $\rC$ region modes ranged between 559
at low temperatures to 11338 at the highest temperature studied.

\subsection{Simulation of expanded interference patterns between two systems}\label{siminterference}
In order to make a direct comparison with the experimental results of
\cite{Hadzibabic2006}, we have generated synthetic interference patterns and 
implemented the experimental analysis technique.  Our simulated imaging
geometry is identical to that found in \cite{Hadzibabic2006}, with expansion
occurring in the $z$-direction. The interference pattern is formed in the
$x$-$z$ plane via integration of the density along the $y$-direction
(``absorption imaging'').

 Our algorithm for obtaining the interference pattern due to our classical field is 
very similar to that presented in \cite{Hadzibabic2004}.  Our above cutoff
thermal cloud is taken into account separately.
We consider a pair of fields $\psi_\rC^{(1)}(x,y), \psi_\rC^{(2)}(x,y)$ from different times
during the simulation, chosen such that the fields can be considered
independent.  The 3D wavefunction corresponding to each field is
reconstructed by assuming a harmonic oscillator ground state in the
tight-trapping direction.  These two reconstructed fields are spatially
separated by $\Delta=3$ $\mu$m, corresponding to the period of the optical lattice in
\cite{Hadzibabic2004}.

Given this initial state, we neglect atomic interactions and only account for expansion
in the tightly-trapped direction.  This yields a simple analytical result for 
the full classical field $\psi_\rC(x,y,z,\tau)$ at later times.  The contribution of the
above-cutoff atoms is included by an incoherent addition of intensities.
The result is integrated along the $y$-direction to simulate the effect of
absorption imaging with a laser beam, that is,
\begin{align}
n_{\rm{im}}(x,z)
	&=\int_0^{L'} dy\,\bktsq[\Big]{\abs[\big]{\psi_\rC^{(T)}(x,y,z,\tau)}^2+n_\rI(x,y,z,\tau)}, \\
\psi_\rC^{(T)}
	&=\psi_\rC^{(1)}(x,y,z,\tau)+\psi_\rC^{(2)}(x\!-\!\Delta,y,z,\tau).
\end{align}
Rather than integrate the full field along the $y$-direction, we use only
a slice of length $L' = 10$ $\mu$m in keeping with the experimental geometry of
Ref.~\cite{Hadzibabic2006}.

The interference patterns, $n_{\rm{im}}(x,z)$, generated this way contained fine spatial detail not seen in the experimental images. To make a more useful comparison to experiment it is necessary to account for the finite optical imaging resolution  by applying a Gaussian convolution in the $x$-$z$ plane with standard deviation 3 $\mu$m.
\cite{Hadzibabic2007priv}.

\begin{figure}[htbp]
  \includegraphics[height=6.3cm]{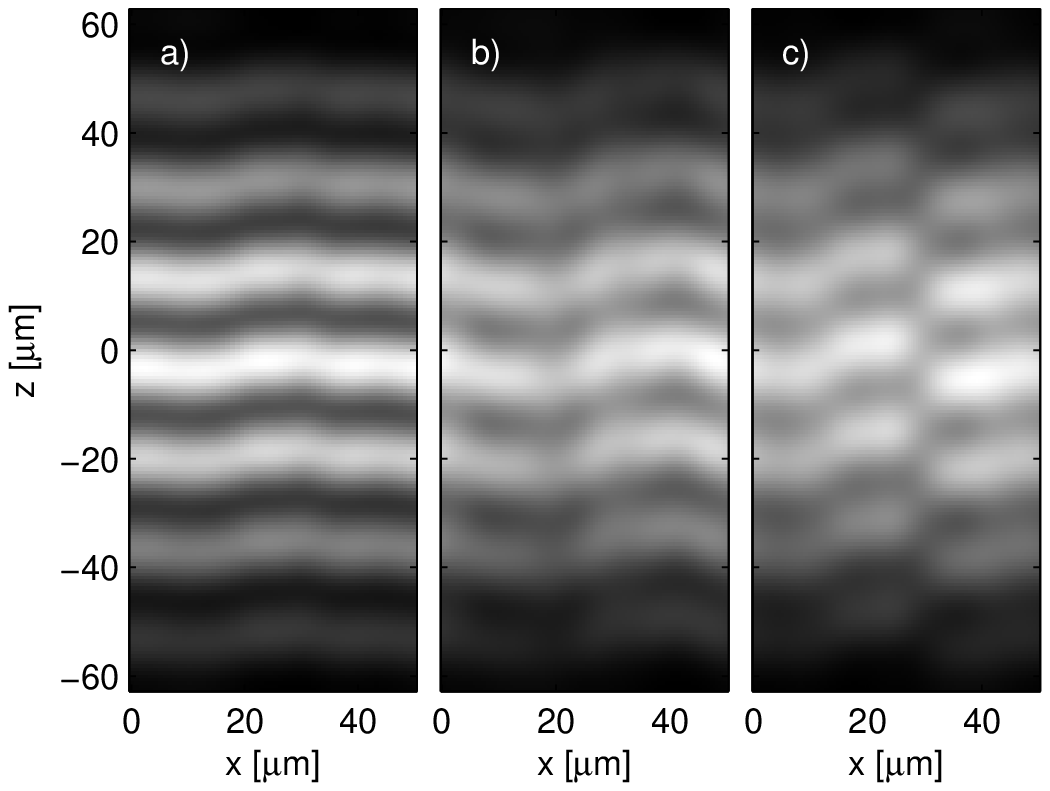}
  \includegraphics[height=6.3cm]{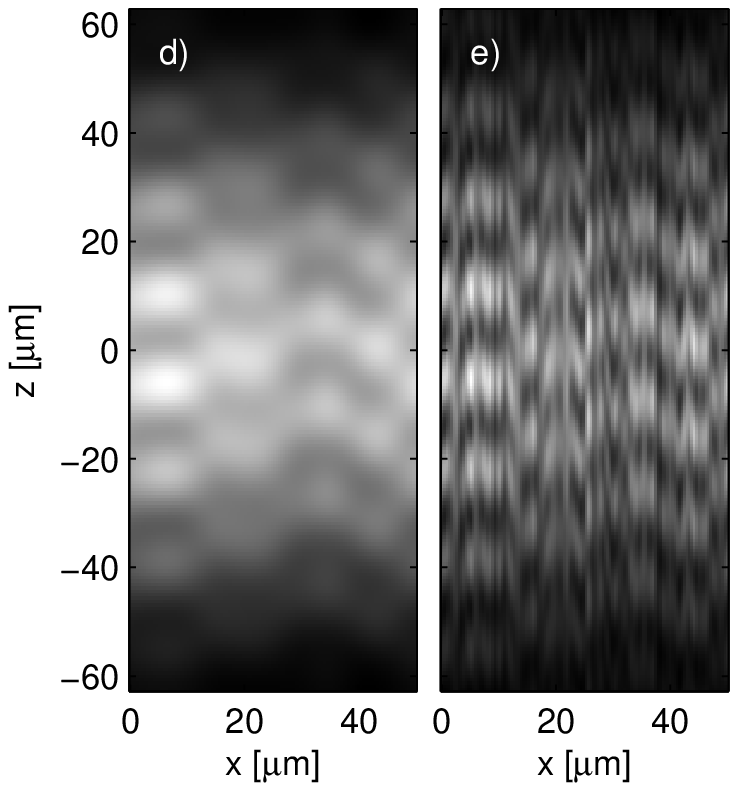}
  \caption{ \label{fig:interference_eg}
  Synthetic interference patterns generated from the 50 $\mu$m grid by simulation of the
  experimental procedure of Ref.~\cite{Hadzibabic2006}.  (a) At low temperatures,
  $T \approx 0.5\Tkt$, the interference fringes are straight.  (b) Just below 
  the transition temperature, $T \approx 0.95 \Tkt$, the fringes become wavy due 
  to decreased spatial phase coherence.  Phase dislocations become common at
  temperatures above the transition, (c) $T \approx 1.05\Tkt$, and (d)
  $T \approx 1.1\Tkt$.  These ``zipper patterns'' indicate the presence of free
  vortices.  (e) When simulation of the finite imaging resolution is disabled,
  the zipper patterns from the field in sub figure (d) are no longer clearly
  visible; the high-frequency details obscure the phase information without
  providing obvious additional information about the existence of vortex pairs.
  }
\end{figure}

In accordance with the Paris experiment, we use a 22~ms expansion time to generate interference patterns for quantitative analysis (see Sec.~\ref{G1interfere}). To obtain characteristic interference images for display in \cite{Hadzibabic2006}, the experiments used a shorter 11 ms expansion  \cite{Hadzibabic2007priv}.  We
exhibit examples of interference patterns at various temperatures in
Fig.~\ref{fig:interference_eg}, for this shorter expansion time. These images show a striking resemblance to 
the results presented in Ref.~\cite{Hadzibabic2006}.

\subsection{Condensate and superfluid fractions} \label{sec:fc_fs}
For a 2D Bose gas in a box we expect a nonzero condensate fraction due to the
finite spacing of low-energy modes.  A central question is whether we can 
observe a distinction between the crossover due to Bose condensation and that
due to BKT physics.  To address this question we have computed both the 
condensate and superfluid fractions from our dynamical simulations.

The condensate fraction in a homogeneous system is easily identified as the
average fractional occupation of the lowest momentum mode.  This is directly
available from our simulations as a time average of the $\v{k} = \v{0}$ mode of
the classical field,
\begin{equation}
	f_c = \Exval{c_\v{0}^* c_\v{0}^{\vphantom{*}}} / N.
\end{equation}

Extracting the superfluid fraction from dynamical classical field simulations
provides a more difficult challenge.  For this we use linear response theory to 
relate the superfluid fraction to the long wavelength limit of the second order
momentum density correlations.  Details concerning the technique are given in
appendix \ref{sec:superfluid_fraction}.

\begin{figure}[htbp]
  \includegraphics[width=8cm]{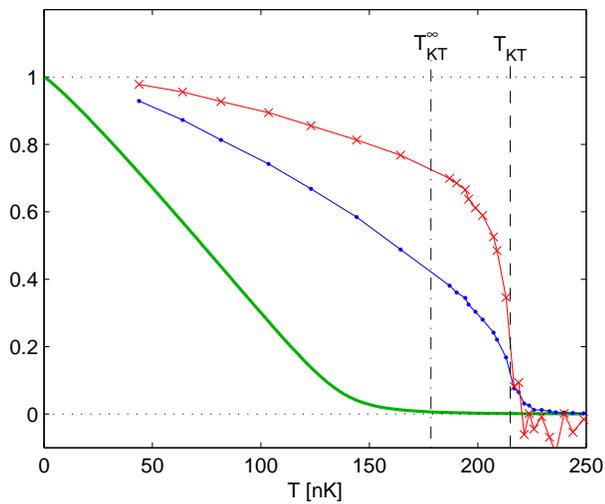}
  \caption{ \label{fig:fc_fs_vs_T}
  (color online).
  Condensate fraction (solid dots) and superfluid fraction (crosses) as 
  functions of temperature for the 100 $\mu$m$^2$ grid.  The transition
  temperature in the thermodynamic limit, $\Tkt^\infty$ \cite{Prokofev2001},
  is shown as a vertical dot-dashed line.  The vertical dashed line shows our
  estimate for the transition temperature in the finite system.  The thick
  solid line is the condensate fraction for an ideal Bose gas in the grand 
  canonical ensemble with the same number of atoms and periodic spatial domain.
  The superfluid fraction becomes negative in places because the extrapolation
  of the momentum correlations to $\v{k} = 0$ is sensitive to statistical noise
  at high temperature (see appendix \ref{sec:superfluid_fraction_numerics} for
  details).
  }
\end{figure}

\begin{figure}[htbp]
  \includegraphics[width=8cm]{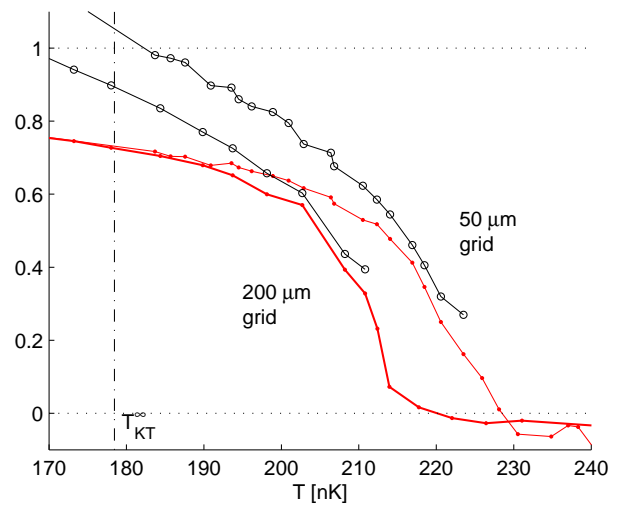}
  \caption{ \label{fig:fs_all_vs_T}
  (color online).
  Detail of the superfluid fraction near the transition temperature.  Solid
  dots represent the calculation based on momentum correlations as
  described in appendix \ref{sec:superfluid_fraction}.  Results for the
  largest and smallest grids are shown (left and right, respectively).  The
  data for the 100 $\mu$m grid is omitted for clarity, but lies between the
  curves shown as expected.  Open circles represent the calculation of
  the superfluid fraction from the associated fitted values for the decay
  coefficient $\alpha$, via Eq.~\eqref{eqn:alpha_rhos_T}.  The open circles 
  terminate where the power law fitting procedure fails.
  }
\end{figure}

Figure \ref{fig:fc_fs_vs_T} compares the results for the superfluid and
condensate fractions computed on the 100 $\mu$m grid.  These results are
qualitatively similar to the results for the larger and smaller grids.  In
particular, we note that there is no apparent separation between temperatures
at which the superfluid and condensate fractions fall to zero.  Also shown in
Fig.~\ref{fig:fc_fs_vs_T} is the condensate fraction for the ideal Bose gas
confined to an identical finite-size box in the grand canonical ensemble.  The large
shift between ideal and computed transition temperatures indicates the
effect of interactions in the 2D system. Because the average system density is uniform, this large shift is to   due to critical fluctuations (also see \cite{Kashurnikov2001a}).

In our calculations we identify the transition temperature, $\Tkt$, 
as where the superfluid fraction falls off most rapidly (i.e., the location of steepest slope on the $\rho_s$ versus $T$ graph; see Fig.~\ref{fig:fc_fs_vs_T}).  As the system
size increases, this transition temperature moves toward the value for an
infinite-sized system, $\Tkt^\infty$ \cite{Prokofev2001}.  This effect is
illustrated by the behavior of the superfluid fraction in
Fig.~\ref{fig:fs_all_vs_T}.

\subsection{First order correlations --- algebraic decay}
\label{sec:g1}
Algebraic decay of the first order correlations, as described by
Eq.~\eqref{eqn:algebraic_decay}, is a characteristic feature of the BKT
phase.  Above the BKT transition, the first order correlations should
revert to the exponential decay expected in a disordered phase.

The normalized first order correlation function, $g^{(1)}$ is defined by
\begin{equation} \label{eqn:g1_definition}
	g^{(1)}(\vec{x},\vec{x}') = \frac{G^{(1)}(\vec{x},\vec{x}')}
		{\sqrt{n(\v{x})n(\v{x}')}},
\end{equation}
where $G^{(1)}(\vec{x},\vec{x}') =
\exval[\big]{\hat{\psi}^\dagger(\vec{x})\hat{\psi}(\vec{x'})}$ is the
unnormalized first order correlation function \cite{Naraschewski1999}.

\subsubsection{Direct calculation of $g^{(1)}$} \label{sec:direct_calc_g1}
In the PGPE formalism the $\rC$ and $\rI$ contributions to the correlation function are additive \cite{Bezett2008}, that is,
\begin{equation}
G^{(1)}(\vec{x},\vec{x}')=G^{(1)}_\rC(\vec{x},\vec{x}')+G^{(1)}_\rI(\vec{x},\vec{x}'),
\end{equation}
where $G^{(1)}_\rC$ and $G^{(1)}_\rI$ are defined in Eqs.~\eqref{Gc1} and \eqref{eqn:GI1}, respectively.
It is interesting to note that $G^{(1)}_\rC$ and $G^{(1)}_\rI$ individually
display an oscillatory decay behavior --- originating from the cutoff --- an
effect which correctly cancels when the two are added together.

\begin{figure}[htbp]
  \includegraphics[width=8cm]{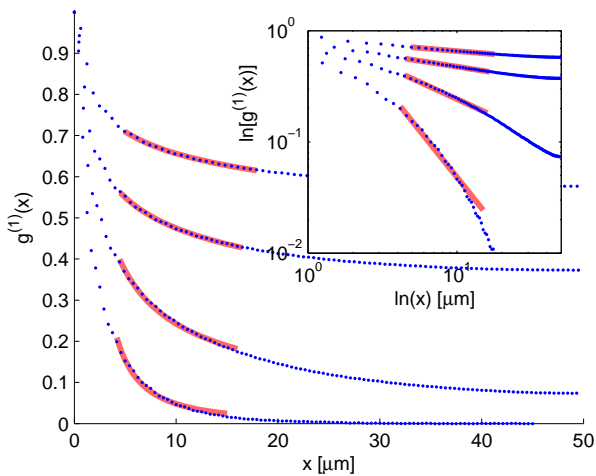}
  \caption{ \label{fig:alpha_fit_eg}
  (color online).
  Sample fits to the algebraic decay of $g^{(1)}$ at various temperatures, 
  ranging from below to above the transition.  High temperatures correspond to
  curves at the bottom of the figure which have rapid falloff of $g^{(1)}$ with
  distance.
  Fits are shown on a log-log scale in the inset to emphasize the failure of a
  power law in describing the behavior of $g^{(1)}$ at high temperature.
  }
\end{figure}

Having calculated $g^{(1)}$, we obtain the coefficient $\alpha$ by fitting the
algebraic decay law, Eq.~\eqref{eqn:algebraic_decay}, using nonlinear least
squares; sample fits are shown in Fig.~\ref{fig:alpha_fit_eg}.  The fit is 
conducted over the region between 10 and 40 de~Broglie wavelengths. The short
length scale cutoff is to avoid the contribution of the non-universal normal
atoms, for which the thermal de~Broglie wavelength sets the appropriate decay
length. The long distance cutoff is chosen to be small compared to the
length scale $L$, to avoid the effect of periodic boundary
conditions on the long range correlations.

\begin{figure}[htbp]
  \includegraphics[width=8cm]{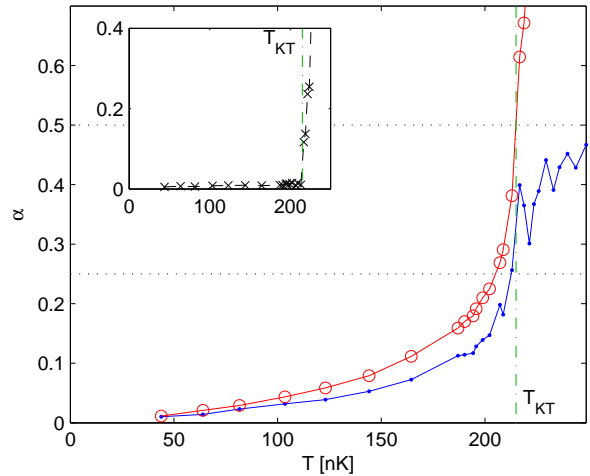}
  \caption{ \label{fig:alpha_comparison}
  (color online).
  Comparison of two methods for determining the algebraic decay coefficient
  $\alpha(T)$ for the first order correlation function $g^{(1)}(\v{x},\v{x}')$.
  The line with circle markers represents direct fits to $g^{(1)}$.  These fits
  fail at the transition temperature as shown by the sharply diverging value 
  of $\abs{\delta(T)}$ in the inset.  The filled points represent the
  values $\alpha'(T)$ obtained from a simulation of the experimental analysis
  procedure of \cite{Hadzibabic2006}, described in Sec.~\ref{G1interfere}.
  Horizontal dotted lines at 0.25 and 0.5 correspond to the expected values
  of $\alpha'$ just below and above the transition, respectively
  \cite{Hadzibabic2006}.  The vertical line is the BKT transition temperature,
  as estimated from the superfluid fraction calculated in Sec.~\ref{sec:fc_fs}.
  }
\end{figure}

The quality of the fitting procedure, and the breakdown of expression \eqref{eqn:algebraic_decay} at the BKT transition can be observed by adding an
additional degree of freedom to the fitting function.  In particular, at each 
temperature we fit the quadratic 
$\ln(g^{(1)}) = A - \tilde{\alpha} \ln(x) + \delta\ln^2(x)$  
and extract the parameter $\delta$ ($\tilde{\alpha} \approx \alpha$ is
discarded).  The abrupt failure of the fits can be observed in the inset of
Fig.~\ref{fig:alpha_comparison} as a sudden increase in the value of
$\abs{\delta(T)}$ --- an effect which is in excellent agreement with the value
of $\Tkt$ as estimated from the superfluid fraction.

\subsubsection{Calculation of $g^{(1)}$ via interference patterns}\label{G1interfere}

So far a direct probe of the \textit{in situ} spatial correlations has not been possible, although important progress has been made by the NIST group \cite{Clade2008}.
In the experiments of Hadzibabic \textit{et al.} \cite{Hadzibabic2006} a scheme proposed by Polkovnikov  \textit{et al.} \cite{Polkovnikov2006} was used to infer these correlations from the ``waviness'' of interference patterns produced by pair of quasi-2D systems (see Sec.~\ref{siminterference}). 
 In this section we simulate the experimental data analysis method, and compare inferred predictions for the correlation function against those we can directly calculate. This allows us to characterize the errors associated with this technique arising from finite size effects and finite expansion time.

To make this analysis we follow the procedure outlined in \cite{Hadzibabic2006}. We fit our numerically generated interference patterns (see Sec.~\ref{siminterference})  to the function 
\begin{equation}
	F(x,z) = G(z) \Bktsq{1 + c(x) \cos\Bkt{\frac{2 \pi z}{D} + \theta(x)}},
\end{equation}
where $G(z)$ is a Gaussian envelope in the $z$-direction, $c(x)$ is the
interference fringe contrast, $D$ is the fringe spacing and $\theta(x)$ is the
phase of the interference pattern in the $z$-direction.

Defining the function
\begin{equation}
	C(L_x) = \frac{1}{L_x} \int_{-L_x/2}^{L_x/2} c(x) e^{i\theta(x)} dx,
\end{equation}
the nature of spatial correlations is then revealed by the manner in which $\exval[\big]{\abs{C(L_x)}^2}$ decays with $L_x$. In particular, we identify the parameter $\alpha'$, defined by $\exval[\big]{\abs{C(L_x)}^2}
\propto L_x^{-2\alpha'}$ \cite{Polkovnikov2006}. For an infinite 2D system in the superfluid regime  ($T<\Tkt^{\infty}$)  $\alpha' = \alpha$ (i.e. $\alpha'$ corresponds to the algebraic decay of correlations). For $T>\Tkt^{\infty}$,
 where correlations decay exponentially, $\alpha'$ is equal to $0.5$.

Fitting $\exval[\big]{\abs{C(L_x)}^2}$ to the algebraic decay law
$A L_x^{\;-2\alpha'}$ we can determine $\alpha'$. A comparison between $\alpha'$
inferred from the interference pattern and $\alpha$ obtained directly from
$g^{(1)}$ is shown in Fig.~\ref{fig:alpha_comparison}.
Both methods give broadly consistent predictions for $\alpha$ when $T<\Tkt$, however our results show that there is a clear quantitative difference between the two schemes, and that $\alpha'$ underestimates the coefficient of algebraic decay in the system (i.e. using $\alpha'$ in Eq.~(\ref{eqn:alpha_rhos_T}) would overestimate the superfluid density). Near and above transition temperature, where the
fits to $g^{(1)}$ fail,  we observe that $\alpha'$ converges toward $0.5$.  The
agreement between $\alpha$ and $\alpha'$ in the low temperature region improves as the size of the grid is increased. 


%

\subsection{Vortices and pairing}
\label{sec:vortices}

The simplest description of the BKT transition is that it occurs as a result of 
vortex pair unbinding: At $T<\Tkt$ vortices only exist in pairs of opposite 
circulation, which unbind at the transition point to produce free vortices that 
destroy the superfluidity of the system.  However, to date there are no direct 
experimental observations of this scenario, and theoretical studies of 2D Bose
gases have been limited to qualitative inspection of the vortex distributions.
In the c-field approach vortices and their dynamics are clearly revealed,
unlike other ensemble-based simulation techniques where the vortices are
obscured by averaging. This gives us a unique opportunity to investigate the
role of vortices and pairing in a dilute Bose gas.

\begin{figure}[htbp]
  \includegraphics[width=7cm]{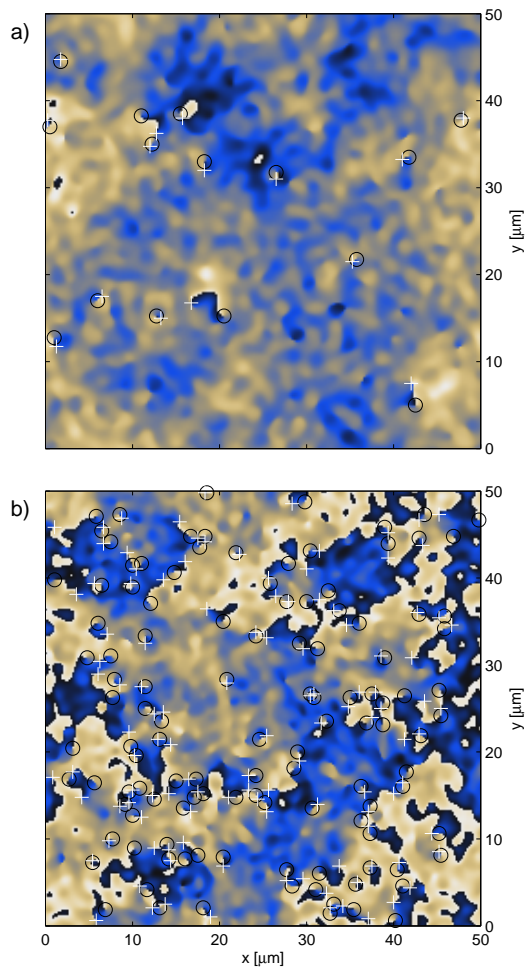}
  \caption{\label{fig:vortex_pairing_eg}
	(color online). Phase profile of a c-field with vortices indicated. 
	 Vortices with clockwise (white $+$) and anticlockwise (black $\circ$) 
	 circulation.  The phase of the classical field is indicated by shading
	 the background between dark blue (phase 0) and light yellow (phase $2\pi$).
	 (a) Distinctive pairing below the transition at $T=207\text{nK}\approx0.93\Tkt$
	 (b) A ``vortex plasma'' above the transition at $T=238\text{nK}\approx1.07\Tkt$.
  }
\end{figure}

We detect vortices in the c-field microstates by analyzing the phase profile of 
the instantaneous field (see appendix \ref{sec:vortex_detection}). An example 
of a phase profile of a field for $T<\Tkt$ is shown in 
Fig.~\ref{fig:vortex_pairing_eg}(a). The vortex locations reveal a pairing 
character, that is, the close proximity of pairs of positive (clockwise) and 
negative (counterclockwise) vortices relative to the average vortex separation.
An important qualitative feature of our observed vortex distributions is that
at high temperatures, pairing does not disappear from the system entirely.
Indeed, most vortices at high temperature could be considered paired or grouped
in some manner, as shown in Fig.~\ref{fig:vortex_pairing_eg}(b).
Perhaps this is not surprising, since positive and negative vortices have a
logarithmic attraction, and we observe them to create and annihilate readily in
the c-field dynamics. However, this does indicate that the use of pairing to
locate the transition may be ambiguous, and we examine this aspect further
below.

\begin{figure}[htbp]
  \includegraphics[width=8cm]{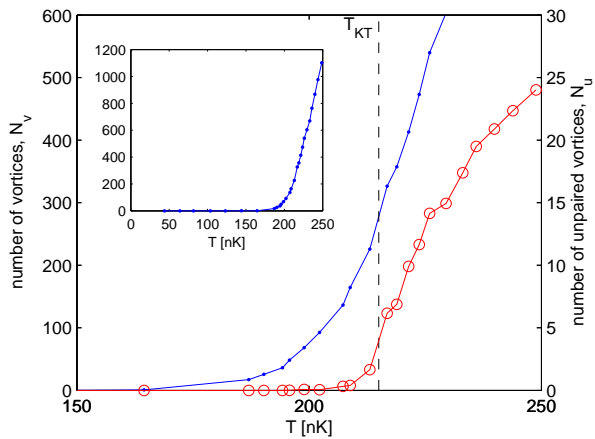}
  \caption{ \label{fig:nv_vs_T}
  Total number of vortices (dots) and number of unpaired vortices (circles) as
  a function of temperature near the transition.  While $N_v$ at the
  transition temperature is already very high, $N_u$ becomes nonzero only close
  to the transition, providing clear evidence of vortex unbinding at work.  The
  inset shows the variation in the total number over the full temperature range
  of the simulations.  Above the transition temperature the growth in the
  number of vortices becomes linear with temperature.
  }
\end{figure}

It is also of interest to measure the number of vortices, $N_v$, present in the
system as a function of temperature (see Fig.~\ref{fig:nv_vs_T}). At the lowest
temperatures the system is in an ordered state, and the energetic cost of
having a vortex is prohibitive. As the temperature increases there is a rapid
growth of vortex population leading up to the transition point followed by 
linear growth above $\Tkt$.

\subsubsection{Radial vortex density}
\begin{figure}[hptb]
  \centering
  \includegraphics[width=8cm]{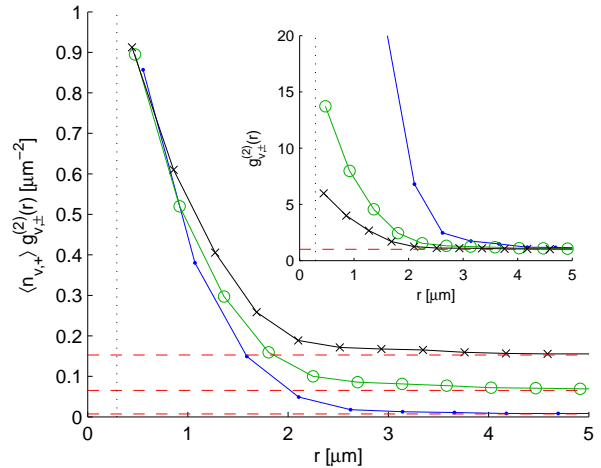}
  \caption{\label{fig:pairing_pdfs}
  (color online).
  Angular average of the two-vortex pair distribution functions for vortices of 
  opposite sign.  Three temperatures centered about the transition are shown:
  dot markers $T=194$nK $\approx 0.9\Tkt$, $f_c = 0.34$; circle markers
  $T=217$nK $\approx 1.01\Tkt$, $f_c = 0.076$; cross markers $T=236$nK $\approx
  1.1\Tkt$, $f_c = 0.006$.  The vertical dotted line shows the value of the
  healing length at $T = 0$.  The main plot shows $g_{v,\pm}^{(2)}$ normalized
  by the positive vortex density; comparable magnitudes for the peaks near 
  $r = 0$ show that vortex pairing remains important over the range of 
  temperatures studied, not only below the transition.  The inset shows
  $g_{v,\pm}^{(2)}$ in the natural dimensionless units for which
  $g_{v,\pm}^{(2)}({r}) \to 1$ as $r \to \infty$.
  }
\end{figure}

The most obvious way to characterize vortex pairing is by defining a pair distribution function for vortices of opposite sign.  Adopting the notation of \cite{Giorgetti2007}, this is
\begin{equation}
G_{v,\pm}^{(2)}(\v{r}) = \Exval{\rho_{v,+}(\v{0}) \rho_{v,-}(\v{r})},
\end{equation}
where $\rho_{v,+}$ is the vortex density function which consists of a sum of
delta-spikes,
\[
\rho_{v,+}(\v{r}) = \sum_{i=1}^{N_{v,+}} \delta(\v{r} - \v{r}^+_i)
\]
for positive vortices at positions $\Bktcl{\v{r}^+_i}$.  We use the analogous
definition for $\rho_{v,-}$.  The associated dimensionless two-vortex
correlation function is
\begin{equation} \label{eqn:g2vpm}
g_{v,\pm}^{(2)}(\v{r}) = \frac{G_{v,\pm}^{(2)}(\v{r})}{\exval{\rho_{v,+}(\v{0})} \exval{\rho_{v,-}(\v{r})}}.
\end{equation}
The angular average of $g_{v,\pm}^{(2)}$ can be calculated directly from
the detected vortex positions using a binning procedure on the pairwise distances
$\norm{\v{r}^+_i - \v{r}^-_j}$, and is shown in Fig.~\ref{fig:pairing_pdfs}.

These results quantify the effect discussed earlier:  Positive and negative vortices show a pairing correlation
which does not disappear above $\Tkt$. The characteristic size of this
correlation, given by twice the width of the peak feature in
Fig.~\ref{fig:pairing_pdfs}, is $l_{\rm{cor}}\sim3\mu$m (taking full width half
maximum).

The shape of our pairing peak is qualitatively similar to that described in
\cite{Giorgetti2007}.  However, in contrast to their results the width does not
appear to change appreciably with temperature.  Additional simulations show
that increasing the interaction strength causes the peak to become squarer and
wider.  It is clear that while the pair size and strength revealed in
$g_{v,\pm}^{(2)}({r})$ does not change appreciably as the transition is
crossed, the amount of pairing relative to the background uncorrelated vortices
changes considerably.  This background of uncorrelated vortices is given by the
horizontal plateau $g_{v,\pm}^{(2)}({r}) \to 1$ at large $r$ as shown in the
inset.

\subsubsection{Revealing unpaired vortices with coarse-graining}


The function $G_{v,\pm}^{(2)}(r)$ clearly indicates the existence of vortex
pairing in the system.  However, it does not provide a convenient way to locate
the transition temperature, since a large amount of pairing exists both below
and above the transition:  The expected number of neighbors for any given
vortex --- roughly, the area of the pairing peak of
$\exval{n_{v,+}}G_{v,\pm}^{(2)}(r)$ shown in Fig.~\ref{fig:pairing_pdfs} ---
does not change dramatically across the transition.  $\exval{n_{v,+}} =
\exval{n_v}/2$ is the expected density of positive vortices.

We desire a quantitative observation of vortex unbinding at the transition and 
have therefore investigated several measures of vortex pairing \footnote{For
example, the Hausdorff distance (see, e.g., \cite{Papadopoulos2005}) between
the set $\{\v{r}^+_i\}$ of positive vortices and the set $\{\v{r}^-_i\}$ of
negative vortices.}.  However, measures based directly on the full set of
vortex positions seem to suffer from the proliferation of vortices at high
temperature --- an effect which tends to wash out clear signs of vortex
unbinding.  With this in mind, we have developed a procedure for measuring the
number of \emph{unpaired} vortices in our simulations, starting from the
classical field rather than the full set of vortex positions.

\begin{figure}[htbp]
  \includegraphics[width=8cm]{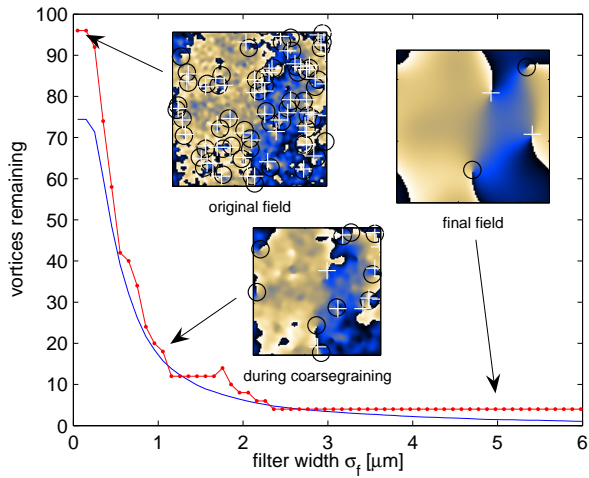}
  \caption{ \label{fig:coarsegrain}
  (color online).
  The coarse-graining procedure: number of vortices as a function of filter
  width for a temperature near the transition.  The smooth curve is an average
  over many realizations of the field, whereas the stepped curve shows typical
  behavior of the number for a single field.  
  Insets show the coarse-grained fields for various filter widths; the
  transformation removes vortex-antivortex pairs which are separated by
  approximately less than the standard deviation of the filter.  In this 
  example $N_u = 4$ unpaired vortices remain at $\sigma_f = 5$ $\mu$m.
  }
\end{figure}

The basis of our approach for detecting unpairing is to
coarse-grain the classical field by convolution with a Gaussian filter of
spatial width (standard deviation) $\sigma_f$.  This removes all vortex pairs
on length scales smaller than than $\sigma_f$.  Figure \ref{fig:coarsegrain}
shows the count of remaining vortices as a function of filter width, along with
some examples of coarse-grained fields. For $\sigma_f\gtrsim l_{\rm{cor}}$, the
number of remaining vortices levels off and only decreases slowly with
increasing $\sigma_f$.  Ultimately the number of remaining vortices goes to
zero as $\sigma_f\to L$.

Setting the filter width to be  larger than the characteristic pairing
distance, $l_{\rm{cor}}$, yields a coarse-grained field from which the pairs
have been removed, but unpaired vortices remain.  In our simulations we have
$l_{\rm{cor}}\approx3$ $\mu$m; we take the vortices which remain after
coarse-graining with a Gaussian of standard deviation $\sigma_f=5$ $\mu$m to
give an estimate of the number of unpaired vortices, $N_u$.  Figure
\ref{fig:nv_vs_T} shows that $N_u$ becomes nonzero only near the transition, in 
contrast to $N_v$ which is nonzero well below $\Tkt$.  The sharp increase in
$N_u$ at $\Tkt$ is a quantitative demonstration of vortex unbinding at work.

In the experiment of Ref.~\cite{Hadzibabic2006}, the fraction of interference 
patterns with dislocations (e.g., see Figs.~\ref{fig:interference_eg}(c) and (d))
was measured.  While isolated vortices are clearly identified by interference
pattern dislocations, a lack of spatial resolution in experiments means that
this type of detection method obscures the observation of tightly bound vortex
pairs.  The experimental resolution of 3 $\mu$m is broadly consistent with the
scale of the coarse-graining filter (i.e., $\sigma_f = 5$ $\mu$m).  With this
in mind, we introduce the quantity $\punpair(T)$, defined as the probability
of observing an unpaired vortex in a $50 \times 50$ $\mu$m control volume at a
given temperature \footnote{We choose a fixed control volume with $L=50$
$\mu$m in order to compare results between simulations with different grid
sizes.}.  For the 50 $\mu$m grid we have simply $\punpair(T) = \Pr(N_u \ge 1)$.

\begin{figure}[htbp]
  \includegraphics[width=8cm]{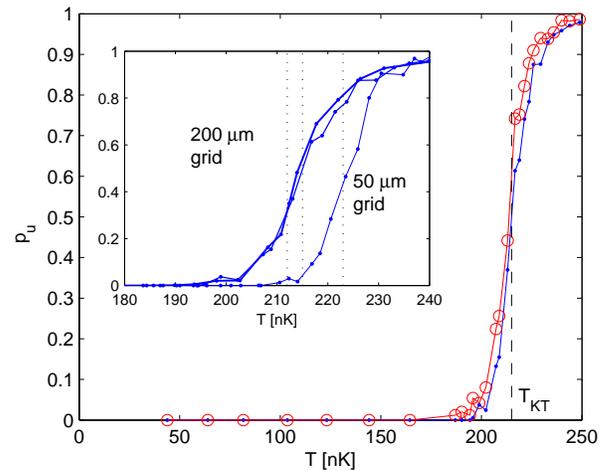}
  \caption{ \label{fig:pairing2}
  (color online).
  Comparison of vortex unpairing measures.  The dots are our pairing measure
  based on coarse-graining the field.  Circles represent the pairing as
  determined by the number of dislocations in the simulated interference
  patterns.  This was the same
  method used in the experimental analysis of \cite{Hadzibabic2006} and
  coincides remarkably well with our coarse-graining based measure.  Both curves are consistent 
  with the vertical line showing the transition temperature $\Tkt$ as
  determined from the superfluid fraction calculation in Sec.~\ref{sec:fc_fs}.
  The inset shows the calculated coarse-grained pairing measure for all three 
  grid sizes, along with vertical lines showing the estimates for $\Tkt$
  derived from the superfluid fraction calculations.
  }
\end{figure}

Computing $\punpair(T)$ from our simulations yields the results shown in
Fig.~\ref{fig:pairing2}.  Our results show a dramatic jump in $\punpair$
at a temperature that is consistent with the transition temperature $\Tkt$
determined from the superfluid fraction calculation presented in
Sec.~\ref{sec:fc_fs}.

From the definition, we expect that $\punpair$ should be close to the
experimentally measured frequency of dislocations.  To demonstrate this
relationship, we have simulated interference patterns (as described in
Sec.~\ref{siminterference}) and detected dislocations using the experimental
procedure of Ref.~\cite{Hadzibabic2006}: A phase gradient $d\theta/dx$ was
considered to mark a dislocation whenever $\Abs{d\theta/dx} > \pi/4$ rad$/\mu$m.
From this we can compute the probability of detecting at least one dislocation
as a function of temperature.  As shown in Fig.~\ref{fig:pairing2}, the results
of this procedure compare very favorably with our measure of pairing based on
$\punpair$.  We note that inhomogeneous effects in experiments probably
broaden the jump in $\punpair$ appreciably compared to our homogeneous
results.

\section{Conclusion}
\label{sec:conclusion}

In this paper we have used c-field simulations of a finite-sized homogeneous 
system in order to investigate the physics of the 2D Bose gas in a regime 
corresponding to current experiments.  We have directly computed the condensate 
and superfluid fractions as a function of temperature, and made comparisons to 
the superfluid fraction inferred by the first order correlation function, and 
using the interference scheme used in experiments. Our results for these 
quantities provide a quantitative test of the interference scheme for a finite
system. 

A beautiful possibility  is the direct experimental observation of vortex-antivortex pairs, their distribution in the system, and hence a quantitative measurement of their unbinding at the BKT transition. We have calculated the vortex correlation function across the transition and provided a coarse-graining scheme for distinguishing unpaired vortices. These results suggest that the dislocations observed in experiments, due to limited optical resolution, provide an accurate measure of the unpaired vortex population and accordingly are a strong indicator of the BKT transition.

We briefly discuss the effect that harmonic confinement (present in experiments) would have on our predictions. The spatial inhomogeneity  will cause the superfluid transition to be gradual, occurring first at the trap center where the density is highest, in contrast to our results where the transition occurs in the bulk.  So far the superfluid fraction for the trapped system has been determined by using the universality result for the critical density in the homogeneous gas \cite{Prokofev2002} in combination with the local density approximation \cite{Holzmann2008, Bisset2009}.  It would be interesting to be able to compute the superfluid fraction independently as we have done here; however it is not clear how to do so.

Bisset \textit{et al.}\ \cite{Bisset2009} used an extension of the c-field method for the trapped 2D gas to examine $g^{(1)}$ and found similar results for the onset of algebraic decay of correlations at the transition.  Their analysis was restricted to the small region near the trap center where the density is approximately constant; we expect the results of our vortex correlation function and the coarse-graining scheme should similarly be applicable to the trapped system in the central region.  Except in very weak traps, the size of this region is relatively small and will likely prove challenging to measure experimentally.

Our results for the homogeneous gas emphasize the clarity with which \textit{ab initio} theoretical methods can calculate quantities directly observable in experiments, such as interference patterns. This should allow direct comparisons  with experiments, providing stringent tests of many-body theory.

\begin{acknowledgments}
The authors are grateful for several useful discussions with Keith Burnett and Zoran Hadzibabic.  CJF and MJD acknowledge financial support from the Australian Research Council Centre of Excellence for Quantum-Atom Optics.
PBB is supported by FRST contract NERF-UOOX0703.
\end{acknowledgments}
\appendix

\section{Simulation using the PGPE} \label{sec:sim_details}

Here we outline our procedure for determining the properties of 
the $\rC$ region and the steps used to create initial states for the PGPE
solver.  The $\rC$ region itself is characterized by the cutoff momentum $K$,
while the initial states are characterized by the energy $E_\rC$ and number
$N_\rC$.  We want to obtain values of these three properties which are 
consistent with a specified temperature $T$ and total number of atoms $N$.



\subsection{Hartree-Fock-Bogoliubov analysis} \label{sec:HFB}
%


To generate an initial estimate of the $\rC$ region parameters we
solve the self-consistent Hartree-Fock-Bogoliubov (HFB) equations in the 
so-called Popov approximation \cite{Griffin1996} to find an approximate thermal
state for the system at a temperature $T$.  The resulting state is a Bose
Einstein distribution of quasiparticles interacting only via the mean-field,
expressed in terms of the quasiparticle amplitudes $u_{\v{k}}$ and $v_{\v{k}}$.

Occupations for the $\rC$ region field may be computed directly from the
quasiparticle occupations via
\begin{equation}
	n_{\v{k}} = \Bkt{u_{\v{k}}^2 + v_{\v{k}}^2} N_B(E_\v{k}) + v_{\v{k}}^2,
\end{equation}
where $N_B$ is the Bose Einstein distribution and $E_\v{k}$ is the
quasiparticle energy which is obtained by solving the Bogoliubov-de~Gennes
equations self consistently~ \cite{Griffin1996}.
This allows us to compute the cutoff as
the maximum value of $\norm{\v{k}}$ consistent with sufficient modal
occupation:
\begin{equation}
K = \max \bktcl{\norm{\v{k}} \colon n_{\v{k}} \ge n_{\text{cut}} }.
\end{equation}
We choose $n_{\text{cut}} = 5$ for
the sufficient occupation condition on the $\rC$ region modes.

The number of atoms below the cutoff may be computed directly from the sum of
the condensate number $N_0$ and the number of $\rC$ region excited state atoms,
$N_{1\rC}$:
\begin{align}
N_\rC = N_0 + N_{1\rC},\quad\text{where}\quad
N_{1\rC} = \sum_{\v{k} \in \rC \backslash \{\v{0}\}} n_{\v{k}}.
\end{align}

For the total energy below the cutoff, we use the expression
\begin{multline}
E_\rC = \frac{\hbar^2}{mL^2} \Bkt{ \frac{g N_0^2}{2} + \lambda N_{1\rC} - g N_{1\rC}^2 } \\
        + \sum_{\v{k} \in \rC \backslash \{\v{0}\}} E_{\v{k}} \Bktsq{N_B(E_{\v{k}}) - v_{\v{k}}^2}
\end{multline}
where $\lambda = g(N_0 + 2 N_{1\rC})$.  Rearranging, this is
\begin{align} \label{EC_hfb}
E_\rC = \frac{\hbar^2 g}{2mL^2}\Bkt{N_\rC^2 + N_{1\rC}^2}
    + \sum_{\v{k} \in \rC \backslash \{\v{0}\}} E_\v{k} \Bktsq{N_B(E_\v{k}) - v_{\v{k}}^2}.
\end{align}
The expression \eqref{EC_hfb} differs from  Eq.~(22) of \cite{Griffin1996} as
we have retained the zeroth order (constant) terms that are required to match
the energy scale of the HFB analysis to the zero point of energy in the
classical field simulations.

\subsection{Initial conditions for fixed total number}


A simple comparison between simulations at varying temperatures can only be
carried out if the total number of atoms is fixed.  This presents a problem in 
our simulations: although the number of atoms and energy of the $\rC$ region
can be directly specified (see Sec.~\ref{sec:initial_given_Ec_Nc}), we may only
determine the total number after performing a simulation.  This is because the
number of atoms in the $\rI$ region depends on the temperature and chemical
potential which are calculated by ergodic averaging of the $\rC$ region
simulations.

Formally, this may be stated as a root-finding problem: solve
\begin{equation}
	N(N_\rC, E_\rC) = N_\text{tot}
\end{equation}
with initial guess provided by the solution to the HFB analysis in
Sec.~\ref{sec:HFB}.  Although both $N_\rC$ and $E_\rC$ affect the total number
$N$, we choose to fix $N_\rC$ to the initial guess and to vary $E_\rC$ until the
desired total number is found.

We note that evaluating the function $N(N_\rC, E_\rC)$ is very computationally
expensive and difficult to fully automate since it involves a simulation and
several steps of analysis.  For this reason we use a nonstandard root finding
procedure:  For the first iteration we simulate three energies about the
initial guess $E_\rC$ such that the results crudely span $N_\text{tot}$; these
three simulations can be performed in parallel which significantly reduces the
time to a solution.  A second guess was obtained by quadratic fitting of
$E_\rC$ as a function of $N$ which gives $N$ accurate to within about 5\% of
$N_\text{tot}$.  An addition iteration using the same interpolation method
takes $N$ to within 0.3\%, which we consider sufficient.

We note that changing $E_\rC$ during the root finding procedure means we have
no direct control over the final temperature of each specific simulation.  In
our case this is not a problem because we only require a range of temperatures
spanning the transition.  In principle one could solve for a given temperature
by allowing $N_\rC$ to vary in addition to $E_\rC$.

\subsection{Initial conditions for given $E_\rC$ and $N_\rC$} \label{sec:initial_given_Ec_Nc}


We compute initial conditions for the $\rC$ region field in a similar way to
\cite{Davis2002}.  Using the representation for the $\rC$ region given by
Eq.~\eqref{eqn:Cfield}, the task is to choose appropriate values for the
$\bktcl{c_\v{n}}$.  As a first approximation, choose the smallest value for a
momentum cutoff $K'$ such that the field with coefficients
\begin{equation} \label{eqn:pgpe_initial1}
c_{\v{n}}
= \begin{cases}
    A e^{i\theta_{\v{n}}} \qquad &\text{for } 0 < \norm{\v{k}} \leq K', \\
    0   \qquad &\text{for } \abs{\v{k}} > K',
  \end{cases}
\end{equation}
has energy greater than $E_\rC$.  Here $A$ is chosen so that the field has
normalization corresponding to $N_\rC$ atoms, and $\theta_{\v{n}}$ is a
randomly chosen phase which is fixed for each mode at the start of the
procedure.  The random phases allow us to generate many unique random initial
states at the same energy.

By definition, the field defined by \eqref{eqn:pgpe_initial1} has energy
slightly above the desired energy.  This problem is solved by mixing it with
the lowest energy state:
\begin{equation}
c_{\v{n}}
= \begin{cases}
    A' e^{i\theta_{\v{0}}} \qquad &\text{for } \v{n} = \v{0}, \\
    0   \qquad &\text{elsewhere},
  \end{cases}
\end{equation}
using a root finding procedure to converge on the desired energy $E_\rC$.
The scheme generates random realizations of a non-equilibrium
field with given $E_\rC$ and $N_\rC$ which are then simulated to equilibrium
before using ergodic averaging for computing statistics.

\section{$\rI$ region integrals} \label{sec:above_cutoff_integrals}

Our assumed self-consistent Wigner function (Sec.~\ref{sec:above_cutoff}) for
the $\rI$ region atoms takes a particularly simple form in the homogeneous
case:
\begin{equation}
W(\v{k},\v{x})
  = \frac{1}{(2\pi)^2} \frac{1}{e^{(\hbar^2\v{k}^2/2m + 2\hbar^2 gn_\rC/m - \mu_\rC)/k_B T} - 1 }.
\end{equation}

The above-cutoff density may then be found by direct integration:
\begin{align}
n_\rI(\v{x}) &= \int_{\norm{\v{k}} \ge K} d^2\v{k} \; W_\rI(\v{k},\v{x}), \\
           &= -\frac{1}{\lambda^2} \ln \Bktsq{1 - e^{-(\hbar^2K^2/2m + 2\hbar^2 gn_\rC/m - \mu_\rC)/k_B T}},
\end{align}
with $\lambda$ the thermal de Broglie wavelength.

In a similar way, the assumed Wigner function allows any
desired physical quantity to be estimated via a suitable integral.  A particular
quantity of interest in the current work is the first order correlation
function,  which can be obtained from the Wigner function as \cite{Naraschewski1999}
\begin{equation}
  G^{(1)}_\rI(\v{x}, \v{x}') = \int_{\norm{\v{k}} \ge K} d^2\v{k}\; e^{-i\v{k}\cdot(\v{x}-\v{x}')} \; W_\rI\bkt[\big]{\v{k},\tfrac{\v{x}+\v{x}'}{2}}.
\end{equation}

This integral is of the general form
\begin{equation}
  I_1(\v{r}) := \int_{\norm{\v{k}} > K} d^2\v{k}\; \frac{e^{-i\v{k}\cdot\v{r}}}{e^{A\v{k}^2 + B} - 1},
\end{equation}
for constants $A$ and $B$.  Noting that $I_1$ depends only on the length,
$r$ of $\norm{\v{r}}$, and transforming $k$ to polar coordinates $(\kappa, \theta)$, 
we have
\begin{align}
  I_1(\v{r}) &= \int_{K}^\infty d\kappa  \frac{\kappa}{e^{A\kappa^2 + B} - 1} \int_0^{2\pi} d\theta\; e^{-ir\kappa\cos\theta},\\
  &=  \int_{K}^\infty d\kappa  \frac{\kappa}{e^{A\kappa^2 + B} - 1} \,2 \Bktsq{\Gamma(\tfrac{1}{2})}^2 J_0(r\kappa),
\end{align}
(see \cite[p902]{Gradshteyn2000} for the Bessel function identity).

Thus we obtain $G_\rI^{(1)}(\v{x}, \v{x}')$ in terms of 
a one dimensional integral which may be performed numerically:
\begin{equation}
  G_\rI^{(1)}(\v{x}, \v{x}')
  = \frac{1}{2\pi} \int_{K}^\infty  d\kappa\;
    \frac{\kappa J_0(\kappa\norm{\v{x}-\v{x}'})}{e^{(\hbar^2 \kappa^2/2m + 2\hbar^2 gn_\rC/m - \mu_\rC)/k_B T} - 1}.
  \label{eqn:GI1}
\end{equation}

\section{Vortex detection} \label{sec:vortex_detection}
The defining feature of a ``charge-$m$'' vortex is that the phase $\theta$ of
the complex field $\psi(\v{x}) = \abs{\psi{(\v{x})}} e^{i\theta(\v{x})}$
changes continuously from $0$ to $2m\pi$ around any closed curve which circles
the vortex core.  We express our field $\psi$ on a discrete grid in position
space; the aim of vortex detection is then to determine which grid plaquettes
(that is, sets of four adjacent grid points) contain vortex cores.

To obtain the phase winding about a plaquette, first consider the phase at two
neighboring grid points A and B.  We are interested in the unwrapped phase
difference $\Delta\theta_{\text{AB}}$ between the grid points; unwrapping
ensures that the phase is \emph{continuous} between A and B.  (In the discrete
setting such continuity is poorly defined; the best we can do is to correct for
the possibility of $2\pi$ phase jumps by adding or subtracting factors of
$2\pi$ so that $\Abs{\Delta\theta_{\text{AB}}} < \pi$.)  The unwrapped phase
differences around a grid plaquette tell us a total phase change $\theta_\text{wrap}
= \sum_i \Delta\theta_{i,i+1} = 2m\pi$ where $m\in\mathbb{Z}$ is the winding
number or ``topological charge''.

Due to the necessity of unwrapping the phase, a four-point grid plaquette cannot 
unambiguously support vortices with charge larger than one.  Luckily, such 
vortices are energetically unfavorable in 2D Bose gases
\cite[p.83]{Pitaevskii_Stringari2003} so we need only concern ourselves with
detecting vortices with winding number $\pm1$ in this work.  The positions
obtained from a given run of our vortex detection algorithm are the labeled
$\{\v{r}^+_i\}$ and $\{\v{r}^-_i\}$ for winding numbers $+1$ and $-1$,
respectively.

\section{Superfluid fraction} \label{sec:superfluid_fraction}

One of the important characteristics of the BKT transition is the presence of
superfluidity, even in the absence of conventional long range order.  In the 
following we describe a method to calculate the superfluid fraction from our
classical field description.  The method is attractive because it makes use of 
momentum correlations which may be extracted directly from our equilibrium
simulations without any need to introduce additional boundary conditions or
moving defects.

\subsection{Superfluid fraction via momentum density correlations}

Our derivation is based on the procedure presented in
\cite[p.214]{Forster1975}, (see also \cite{Baym1968} and
\cite[p.96]{Pitaevskii_Stringari2003}).
The central idea is to establish a relationship between i) the
autocorrelations of the momentum density in the simulated ensemble and ii) the
linear response of the fluid to slowly moving boundaries; (i) is a quantity we
can calculate, while (ii) is related to the basic properties of a superfluid
via a simple thought experiment.

\begin{figure}[hptb]
	\centering
	\psfrag{velu}[][]{$\v{u}$}
	\psfrag{limy}[][]{$\displaystyle\lim_{L_y\to\infty}$}
	\psfrag{limx}[][]{$\displaystyle\lim_{L_x\to\infty}$}
	\psfrag{superflui}[][]{Superfluid}
	\psfrag{Ly}[][]{$L_y$}
	\psfrag{Lx}[][]{$L_x$}
	\psfrag{aa}[][]{a)}
	\psfrag{bb}[][]{b)}
	\psfrag{cc}[][]{c)}
	\psfrag{dd}[][]{d)}
	\includegraphics[width=8cm]{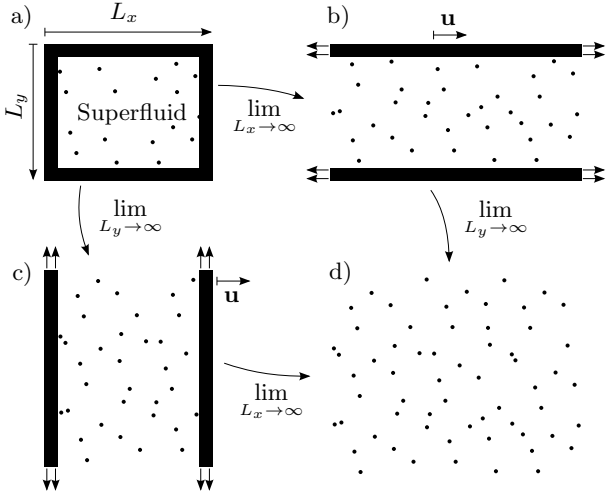}
	\caption{ \label{fig:superfluid_thought_expt}
	Thought experiment used in deriving the superfluid density.  The walls
	move with velocity $\v{u}$ in the $x$ direction.  To begin with, we imagine
	that the superfluid sits in a box of dimensions $L_x \times L_y$ as shown
	in a).  We later take the limit as the box walls recede to infinity to get 
	the thermodynamic limit d).  The order of the limits is critically 
	important: the path b) leads to superflow while the path c) results in the 
	entire fluid moving along with the walls.
	}
\end{figure}

To connect the macroscopic, phenomenological description of superfluidity with
our microscopic theory, we make use of the thought experiment shown 
schematically in Fig.~\ref{fig:superfluid_thought_expt}(b): Consider an 
infinitely long box, $B$ containing superfluid, and accelerate the box along
its long axis until it reaches a small velocity $\v{u}$.  Due to viscous
interactions with the walls, such a box filled with a normal fluid should have
a momentum density at equilibrium of $\exval{\md}_\v{u} = n \v{u}$.  The
notation $\exval{\cdot}_\v{u}$ denotes an expectation value in the ensemble
with walls moving with velocity $\v{u}$.

Because superfluids are nonviscous, the observed value for the momentum 
density in a superfluid is less than the value $n\v{u}$ expected for a
classical fluid.  In Landau's two-fluid model we attribute the observed
momentum density, $\rho_n \v{u}$, to the ``normal fraction'' where $\rho_n$ is
the normal fluid density.  The superfluid fraction remains stationary in the
lab frame, even at equilibrium and makes up the remaining mass with density
$\rho_s = n - \rho_n$.

In order to apply the usual procedures of statistical mechanics to the thought
experiment, we consider two frames: the ``lab frame'' in which the walls move
with velocity $\v{u}$ in the $x$-direction and the ``wall frame'' in which the
walls are at rest.

Assuming that the fluid is in thermal equilibrium with the walls, the density matrix in
the grand canonical ensemble is given by the usual expression $\hat{\rho} =
e^{-\beta(H_{\v{u}} - \mu N)} / \Tr\Bkt{e^{-\beta(H_{\v{u}} - \mu N)}}$ where
$H_{\v{u}}$ is the Hamiltonian of the system in the wall frame and $\beta = 1/k_B T$.
A Galilean transformation relates $H_{\v{u}}$ to the Hamiltonian in the lab
frame, $H_{\v{u}} = H - \v{u}\cdot \hat{\v{P}}+ \frac{1}{2}M u^2$, where $\hat{\v{P}} =
\int_B \md(\v{x}) d^2r$ is the total momentum, $M = mN$ is the total mass and
$\md(\v{x})$ is the momentum density operator at point $\v{x}$.  The
expectation value for the momentum density in the presence of moving walls is
then given by the expression
\begin{align}
\Exval{\md(\v{x})}_\v{u} 
  &= \Tr[\hat{\rho}\,\md(\v{x})], \\
  &= \frac{\Tr\bkt[\big]{e^{-\beta(H - \hat{\v{P}}\cdot \v{u} + (m u^2 / 2 - \mu) N)}\md(\v{x})}}
  {\Tr\bkt[\big]{e^{-\beta(H - \hat{\v{P}}\cdot \v{u} + (m u^2 / 2 - \mu) N}}}.
\end{align}
Expanding this expression to first order in $\v{u}$ yields
\begin{equation}
  \Exval{\md(\v{x})}_\v{u} = \Exval{\md(\v{x})} +
  \beta\bkt[\big]{\exval{\md(\v{x}) \hat{\v{P}}\cdot \v{u}}
  - \Exval{\md(\v{x})}\exval{\hat{\v{P}}\cdot \v{u}}},
\end{equation}
where all the expectation values on the right hand side are now taken in the
\emph{equilibrium} ensemble with the walls at rest.  Since $\exval{\md(\v{x})}
= 0$ in our equilibrium ensemble, this simplifies to
\begin{align}
  \Exval{\md(\v{x})}_\v{u} &= \beta\exval[\big]{\md(\v{x}) \hat{\v{P}}} \cdot \v{u}, \\
  &= \beta\int_B d^2\v{x}' \, \Exval{\md(\v{x})\md(\v{x}')} \cdot \v{u}, \label{eqnMomdensRaw}
\end{align}
where $\md(\v{x})\md(\v{x}')$ is a dyad [i.e., a rank-two tensor; the outer 
product of $\md(\v{x})$ and $\md(\v{x}')$].

To make further progress, we wish to take the limit as the system gets very 
large (write ``$B \to \infty$'').  To this end, we first consider some
properties of the correlation functions in the infinite system.  The infinite
system is homogeneous, which implies that $\Exval{\md(\v{x})\md(\v{x}')}_\infty
= \Exval{\md(\v{x}+\v{r})\md(\v{x}'+\v{r})}_\infty$ for any $\v{r}$, where
$\Exval{\cdot}_\infty$ indicates an average in the infinite system.  As
a consequence, we may express the correlations --- in the infinite system ---
in terms of the Fourier transform in the relative coordinate $\v{x}'-\v{x}$:
\begin{align}
\Exval{\md(\v{x})\md(\v{x}')}_\infty
	&= \Exval{\md(\v{0})\md(\v{x}'-\v{x})}_\infty, \\
	&= \frac{1}{(2\pi)^2} \int d^2\v{k} \, e^{i\v{k}\cdot(\v{x}'-\v{x})} \chi(\v{k}),
\end{align}
where all the important features of the correlations are now captured by the tensor
\begin{equation}
  \chi(\v{k}) = \int d^2\v{r} \, e^{-i\v{k}\cdot\v{r}} \Exval{\md(\v{0})\md(\v{r})}_\infty.
  \label{eqn:chiDefinition}
\end{equation}
Because of the isotropy of the fluid in the infinite system, $\chi(\v{k})$
obeys the transformation law $\chi(O\v{k}) = O^{-1}\chi(\v{k})O$ for any
$2\times 2$ orthogonal matrix $O$.  This implies that $\chi$ may be decomposed
into the sum of longitudinal and transverse parts:
\begin{equation}
  \chi(\v{k})
    = \tilde{\v{k}}\tilde{\v{k}} \chi_l(k) + \bkt[\big]{I - \tilde{\v{k}}\tilde{\v{k}}} \chi_t(k)
  \label{eqn:chiDecomp}
\end{equation}
where $\tilde{\v{k}} = \v{k}/k$, $k = \norm{\v{k}}$ and $I$ is the identity.  
The transverse and longitudinal functions $\chi_t$ and $\chi_l$ are scalars
which depend only on the length $k$.

We now return our attention to the finite system.  If the finite box $B$ is
large then the momentum correlations in the bulk will be very similar to the
values for the infinite system.  Therefore, when $\v{x}$ and $\v{x}'$ are far
from the boundaries, we may approximate
\begin{align}
\Exval{\md(\v{x})\md(\v{x}')}
	&\approx \Exval{\md(\v{x})\md(\v{x}')}_\infty \\
	&= \frac{1}{(2\pi)^2} \int d^2\v{k} \, e^{i\v{k}\cdot(\v{x}'-\v{x})} \chi(\v{k})
\end{align}
which in combination with Eq.~\eqref{eqnMomdensRaw} yields
\begin{align}
  \Exval{\md(\v{x})}_\v{u}
    &\approx \beta \int_B d^2\v{x}' \frac{1}{(2\pi)^2} \int d^2\v{k} \, 
      e^{i\v{k}\cdot(\v{x}' - \v{x})} \chi(\v{k}) \cdot \v{u} \\
    &= \beta \int d^2\v{k} \, \Delta_B(\v{k}) e^{i\v{k}\cdot\v{x}} \chi(\v{k}) \cdot \v{u}.
\end{align}
Here we have defined the nascent delta function $\Delta_B(\v{k}) := \frac{1}{(2\pi)^2}
\int_B d^2\v{x}' \, e^{i\v{k}\cdot\v{x}'}$ which has the property
$\Delta_B(\v{k}) \to \delta(\v{k})$ as $B \to \infty$.

We are now in a position to carry out the limiting procedure to increase the
box size to infinity.  However, care must be taken because the simple expression
$\lim_{B\to\infty} \Exval{\md(\v{x})}_\v{u}$ is not well defined without 
further qualification of the limiting process $B\to\infty$.

To resolve this subtlety we must insert a final vital piece of physical reasoning.
Let us assume for simplicity that $\v{u}$ is directed along the $x$-direction,
and the box $B$ is aligned with the $x$ and $y$ axes with dimensions
$L_x{\times}L_y$.  As shown in Fig.~\ref{fig:superfluid_thought_expt}, there
are two possibilities for taking the limits, representing different physical
situations.

On the one hand [Fig.~\ref{fig:superfluid_thought_expt}(b)], we may take the
limit $L_x\to\infty$ first, which gives us an infinitely long channel in which
superfluid can remain stationary while only the normal fraction moves with the
walls in the $x$-direction.  We have
\begin{align}
\rho_n\v{u}
	&= \lim_{L_y\to\infty} \lim_{L_x\to\infty} \Exval{\md(\v{x})}_\v{u} \\
	&= \lim_{L_y\to\infty} \lim_{L_x\to\infty} \beta \int d^2\v{k} \, \Delta_B(\v{k}) e^{i\v{k}\cdot\v{x}} \chi(\v{k}) \cdot \v{u} \\
    &= \beta \lim_{k_y\to 0} \lim_{k_x\to 0} \chi(\v{k}) \cdot \v{u}
\end{align}
where we use the fact that $\Delta_B(\v{k})$ can be decomposed into the product
$\Delta_{L_x}(k_x)\Delta_{L_y}(k_y)$ with $\Delta_{L}(k) \to \delta(k)$ as
$L\to\infty$.  Employing the decomposition of $\chi$ given in
Eq.~\eqref{eqn:chiDecomp} allows the density of the normal fraction to be
related to the transverse component of $\chi$ evaluated at zero:
\begin{equation}
	\rho_n = \beta\lim_{k\to 0}\chi_t(k) = \beta \chi_t(0).
\end{equation}

On the other hand [Fig.~\ref{fig:superfluid_thought_expt}(c)] we may take the
limit $L_y\to\infty$ first, resulting in an infinitely long channel --- with
velocity \emph{perpendicular} to the walls --- in which the entire body of the
fluid must move regardless of the superfluidity.  In a similar way to the 
previous paragraph, $n\v{u} = \beta \lim_{k_x\to \v{0}} \lim_{k_y\to \v{0}}
\chi(\v{k}) \cdot \v{u}$, and making use of the decomposition in
Eq.~\eqref{eqn:chiDecomp}, the total density is related to the longitudinal
component of the correlations:
\begin{equation}
	n = \beta \lim_{k\to 0} \chi_l(k) = \beta \chi_l(0).
\end{equation}

With these expressions, the normal fraction $f_n$ may finally be expressed
directly as
\begin{equation} \label{eqn:normal_fraction}
  f_n = \rho_n/n = \lim_{k\to 0} \chi_t(k) / \lim_{k\to 0} \chi_l(k)
\end{equation}
while the superfluid fraction is $f_s = 1 - f_n$.  Thus, we have expressed the 
superfluid and normal fractions in terms of a correlation function which can be
directly computed from our simulation results.

\subsection{Numerical procedure} \label{sec:superfluid_fraction_numerics}
To determine the superfluid fraction for our system, we need to estimate the 
tensor of momentum density correlations $\chi$ from the simulation results.  
For our finite system constrained to a periodic simulation box, we may compute 
the momentum correlations only at discrete grid points.  The discrete analogue 
of Eq.~\eqref{eqn:chiDefinition} leads to the expression
\begin{equation}
  \chi(\v{k}) \propto \Exval{\v{p}_{\v{k}}\v{p}_{-\v{k}}}
  \label{eqn:chiFinite}
\end{equation}
where the constant of proportionality is not important to the final result, and
$\v{p}_{\v{k}}$ are the discrete Fourier coefficients of $\v{p}(\v{x})$ over 
our simulation box.

The momentum density operator is given by
\begin{equation}
  \md(\v{x}) = \frac{i\hbar}{2}\Bktsq{(\del \hat{\psi}^\dagger(\v{x}) ) \hat{\psi}(\v{x}) - 
  \hat{\psi}^\dagger(\v{x}) \del \hat{\psi}(\v{x}) }
\end{equation}
which may be derived by considering the continuity equation for the number 
density, $\exval[\big]{\hat{\psi}^\dagger(\v{x}) \hat{\psi}(\v{x})}$.  For a 
given classical field Eq.~\eqref{eqn:Cfield}, the Fourier coefficients of 
$\v{p}$ may be written as
\begin{equation}
  \v{p}_{\v{k}} = \frac{\hbar}{2\sqrt{A_B}} \sum_{\v{k}'} (2\v{k}'+\v{k})c^*_{\v{k}'}c_{\v{k}+\v{k}'},
\end{equation}
where $A_B$ is the area of the system.  Computing a value for all $\v{p}_\v{k}$
at each time step, we then evaluate $\chi(\v{k})$ via the usual ergodic
averaging procedure using Eq.~\eqref{eqn:chiFinite}.

Having evaluated $\chi(\v{k})$, we are left with performing the decomposition 
into longitudinal and transverse parts.  For this, simply note that
Eq.~\eqref{eqn:chiDecomp} implies $\chi_l(k) = \tilde{\v{k}}\cdot
\chi(\v{k}) \cdot \tilde{\v{k}}$, and $\chi_t(k) = \tilde{\v{w}}\cdot \chi(\v{k})
\cdot\tilde{\v{w}}$, where $\tilde{\v{w}}$ is a unit vector perpendicular to
$\tilde{\v{k}}$.

\begin{figure}[htbp]
  \includegraphics[width=8cm]{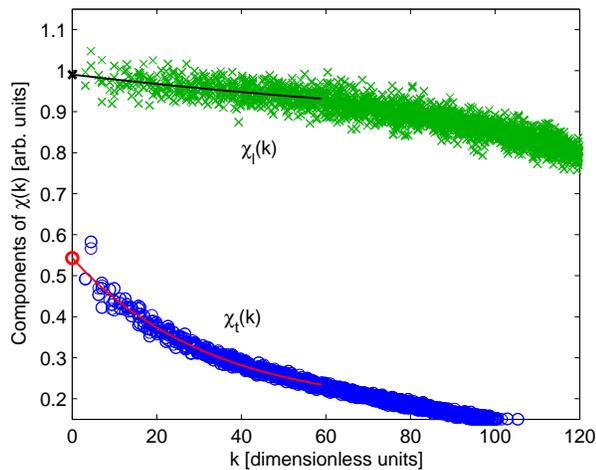}
  \caption{ \label{fig:sfl_lims_eg}
  (color online).
  Example fitting and extrapolation to $k=0$ for the transverse and longitudinal
  components of the momentum density autocorrelation tensor, $\chi$.  The 
  apparent functional form for $\chi_t$ and $\chi_l$ changes with temperature
  --- particularly near the transition --- which along with the sampling noise
  makes them difficult to fit reliably.  The data shown corresponds to a
  temperature slightly below the transition.
  }
\end{figure}

Values for $\chi_t$ and $\chi_l$ may be collected for all angles as a function 
of $k$, and a fitting procedure used to perform the extrapolation $k \to 0$; 
this procedure is illustrated in Fig.~\ref{fig:sfl_lims_eg}.  At low 
temperatures, the extrapolation is quite reliable, but becomes more difficult 
near the transition where sampling noise increases and $\chi_t(k)$ changes
rapidly near $k=0$.  Without a known functional form, we settled for a
quadratic weighted least squares fit of $\ln(\chi_t)$ and $\ln(\chi_l)$ versus $k$.
A weighting of $1/k$ was used to counteract the fact that the density of 
samples of $\chi$ vs $k$ scales proportionally with $k$ due to the square grid 
on which $\chi(\v{k})$ is evaluated.  The logarithm was used to improve the fits
of $\chi_t$ very near the transition where it varies non-quadratically near
$k=0$.  The fitting procedure and extrapolation to $k=0$ generally produces
reasonable results, but is somewhat sensitive to numerical noise.  For this
reason, the computed superfluid fraction at high temperatures is not exactly
zero (see Fig.~\ref{fig:fc_fs_vs_T}).

\end{document}